\documentclass[
    ,final            
  ]
  {aipproc}

\layoutstyle{6x9}

\begin{document}

\newcommand{\mipsmu}{$24\mu$m}
\newcommand{\sfrir}{SFR$_{IR}\geq3M_{\odot}$~yr$^{-1}$}
\newcommand{\magcut}{$-20.5$}

\title{The Importance of AGN in an Assembling Galaxy Cluster}


\author{Kim-Vy H. Tran}{
  address={Mitchell Institute for
Fundamental Physics \& Astronomy, \\ Department of Physics, Texas A\&M University}
  ,altaddress={Institute for Theoretical Physics, University of Z\"urich }
}

\author{ Am\'elie Saintonge}{
  address={Institute for Theoretical Physics, University of Z\"urich }
}

\begin{abstract}

We present results from our multi-wavelength study of SG1120, a super
galaxy group at $z=0.37$ that will merge to form a galaxy cluster
comparable in mass to Coma.  We have spectroscopically confirmed 174
members in the four X-ray luminous groups that make up SG1120, and
these groups have velocity dispersions of
$\sigma_{1D}=303-580$~km~s$^{-1}$.  We find that the supergroup has an
excess of \mipsmu~members relative to CL~1358+62, a rich galaxy
cluster at $z=0.33$.  SG1120 also has an increasing fraction of
\mipsmu~members with decreasing local galaxy density, $i.e.$ an
infrared-density relation, that is not observed in the rich cluster.
We detect nine of the group galaxies in VLA 1.4 Ghz imaging, and
comparison of the radio to total infrared luminosities indicates that
$\sim30$\% of these radio-detected members have AGN.  The radio map
also reveals that one of the brightest group galaxies has radio jets.
We are currently analysing the 1.4 Ghz observations to determine if
AGN can significantly heat the intrahalo medium and if AGN are related
to the excess of \mipsmu~members.

\end{abstract}

\maketitle


\section{Introduction}

Galaxy groups may be the key to understanding the interplay between
galaxy evolution and environment because: 1) most galaxies in the
local universe are in groups \citep[$e.g.$][]{geller:83}; and 2)
hierarchical structure formation predicts that galaxy clusters
assemble from the merger and accretion of smaller structures such as
groups \citep{peebles:70}.  Observational studies also have found that
the quenching of star formation needed to transform an active galaxy
into a passive system occurs outside the cluster cores ($R_p>2$~Mpc)
\citep{gomez:03}, $i.e.$ in lower density (group-like) environments.

Although less massive than galaxy clusters, galaxy groups in the local
universe do have more in common with clusters than with the field
population, $e.g.$ higher early-type fractions and lower mean star
formation rates than the field
\citep{zabludoff:98a,tran:01,rasmussen:08}.  However, whether star
formation in intermediate redshift groups is initially enhanced or
simply quenched relative to the field is debated.  Another closely
linked question is whether active galactic nuclei (in radio/mechanical
mode) have a short but critical role in quenching the star formation
\citep{bower:06,croton:06}, $e.g.$ simulations by
\citet{bhattacharya:08} show that AGN feedback does decrease the gas
density and star formation in galaxies in the group core.  AGN can
also heat the intra-halo medium and thus help explain the high entropy
floors observed in galaxy groups and clusters \citep{ponman:99}.  Note
that several studies find that the AGN fraction in local clusters
increases with decreasing velocity dispersion (mass)
\citep{popesso:06,sivakoff:08}, $i.e.$ the AGN fraction should be
higher in group-massed halos ($M\sim10^{14}M_{\odot}$).

The question then is whether the evolution of galaxies in clusters is
driven primarily on group or on cluster scales.  Our discovery of a
supergroup of galaxies at $z=0.37$ allows us to uniquely answer this
question.  The supergroup (hereafter SG1120) is composed of multiple
galaxy groups that we have shown will merge into a cluster comparable
in mass to Coma by $z\sim0$ \citep{gonzalez:05}.  First results from
our multi-wavelength study show that the group galaxies are in
transition: SG1120 has a high fraction of early-type members
\citep{kautsch:08}, yet several of the most massive group galaxies are
growing by dissipationless merging at $z<0.4$ \citep{tran:08}.  Here
we test whether the group galaxies have enhanced activity relative to
the field by combining MIPS \mipsmu~and VLA 1.4 Ghz imaging to measure
total star formation rates and identify AGN.

\section{Results}

Our multi-wavelength study of SG1120 ($z=0.37$) is unique among
existing surveys at intermediate redshift because we: 1) have a large
number (174) of spectroscopically confirmed group galaxies; 2) compare
to the massive, dynamically relaxed cluster CL1358+62 ($z=0.33$, 232
members) \citep{fisher:98}; and 3) also compare to a sample of field
galaxies (87 galaxies at $0.25\leq z\leq0.45$) that have been observed
and analyzed in the same manner as the group and cluster galaxies
\citep{tran:04a}.  Our \mipsmu~imaging identifies all galaxies with
obscured star formation rates of $3M_{\odot}$~yr$^{-1}$ or greater,
regardless of galaxy mass.

\begin{figure}[!t]
\includegraphics[height=.3\textheight]{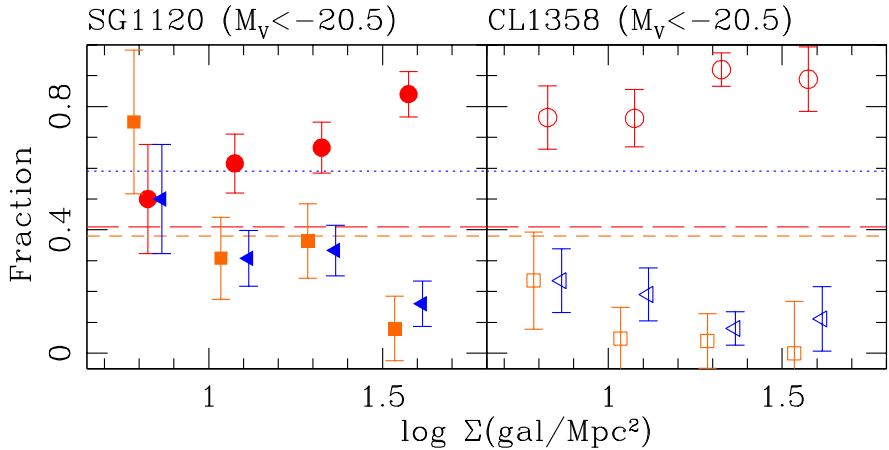}
\caption{Relative fraction of absorption-line (circles), emission-line
(triangles), and \mipsmu~(squares) members as a function of local
galaxy surface density in the groups (SG1120, left panel, filled
symbols) and cluster (CL1358, right panel, open symbols); the points
are offset slightly in $\log\Sigma$ for clarity.  Both panels show the
luminosity-selected ($M_V<$\magcut) samples.  We show the fraction of
absorption-line (long-dash), emission-line (dotted), and
\mipsmu~(short-dash) galaxies in the field as horizontal lines in both
panels.  Only in the supergroup does the \mipsmu~fraction measurably
increase with decreasing local density, $i.e.$ an infrared star
formation--density relation.  At the lowest galaxy densities in the
supergroup, the \mipsmu~fraction is even higher than in the field.
\label{fdensity}}
\end{figure}

Figure~\ref{fdensity} (left panel) shows how the fraction of
\mipsmu~members in the supergroup steadily increases with decreasing
local galaxy density as measured by distance to the $10^{th}$ nearest
spectroscopically confirmed neighbor.  The increasing fraction of
emission-line members with decreasing $\Sigma$ mirrors the trend of
the \mipsmu~members, and the absorption-line fraction changes
accordingly.  In contrast, the \mipsmu~population in the cluster
(right panel) shows a visibly weaker trend with local environment. The
group galaxies have an infrared-density relation where at the lowest
galaxy densities, the \mipsmu~fraction in SG1120 is even higher than
in the field ($\bar{z}=0.35$).  A complete analysis and discussion of
the \mipsmu~population as a function of environment is presented in
\citet{tran:09}.

\begin{figure}[!t]
\includegraphics[height=.45\textheight]{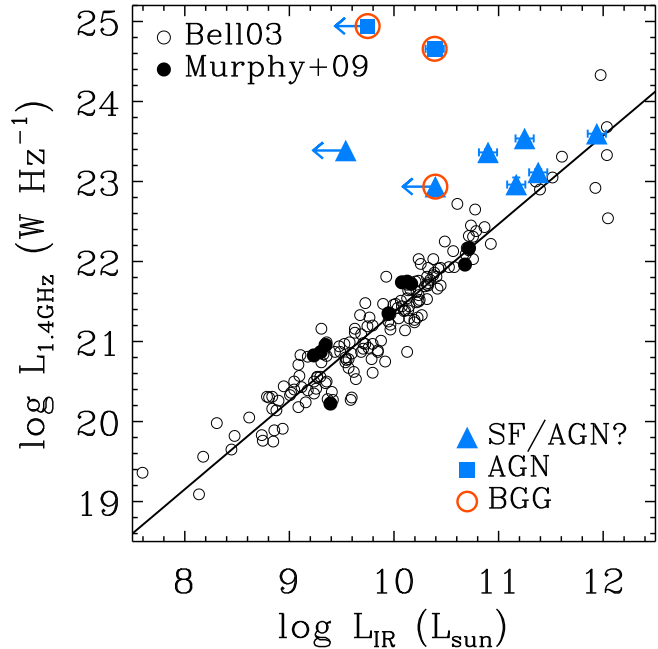}
\caption{Comparison of the total infrared luminosity as measured by
the \mipsmu~flux to the radio luminosity at 1.4~Ghz.  Galaxies
dominated by star formation lie on the solid line, as illustrated by
data on local galaxies (open circles) \citep{bell:03a} and Virgo
members (solid circles) \citep{murphy:09}.  The supergroup galaxies
are shown as solid triangles (star-forming/AGN) and solid squares
(AGN); the radio-detected brightest group galaxies are circled.
\label{tir_radio}} 
\end{figure}

Because the \mipsmu~emission can be due to ongoing star formation or
AGN, we compare the group galaxies' total infrared luminosities to
radio luminosities measured with the VLA in the A-array (resolution of
$\sim1.7''$).  Nine of the group galaxies are detected at 1.4 Ghz and we
show their radio vs. total infrared luminosities in
Fig.~\ref{tir_radio}.  Five of the radio-detected group galaxies have
IR/radio ratios consistent with normal star formation.  However, four
members, including three of the four brightest group galaxies (BGG),
have IR/radio ratios indicative of AGN.  The 1.4 Ghz map also reveals
that one of the BGGs has extended radio jets ($\sim100$ kpc across).

To summarize, we find that SG1120 ($z=0.37$), a protocluster made of
four galaxy groups, has an excess of \mipsmu~members compared to a
massive cluster at $z=0.33$.  However, the fraction of early-type
galaxies in SG1120 is already as high as in the cluster, $i.e.$ the
timescales needed to morphologically transform galaxies into
early-type systems is decoupled from when their star formation is
quenched.  SG1120 also has nine radio sources of which at least
$\sim30$\% are due to AGN rather than ongoing star formation.  We are
currently analyzing the radio observations in detail to better
estimate, $e.g.$ how much energy the AGN inject into the intragroup
medium and whether the AGN are related to the excess of
\mipsmu~sources.


\begin{theacknowledgments}

We thank our collaborators J. Moustakas, A. Gonzalez, L. Bei,
B. Holden, and D. Zaritsky for major contributions to the data
reduction and analysis. Both K.T. and A.S. acknowledge support from
the Swiss National Science Foundation (grant PP002-110576).

\end{theacknowledgments}



\bibliographystyle{aipproc}   

\bibliography{vytran.bib}

\end{document}